\documentclass[11pt]{article}
\usepackage{color, amsmath, amssymb, amsbsy, amsthm, graphicx, bbm, amsfonts, bm, dsfont}
\usepackage{wrapfig}
\usepackage{setspace, booktabs, comment, graphicx, algorithm,  latexsym}
\usepackage[top=1in, bottom=1in, left=1in, right=1in]{geometry}
\usepackage[toc,page]{appendix}
\RequirePackage[numbers]{natbib}
\RequirePackage[colorlinks,allcolors=blue,urlcolor=blue]{hyperref}
\usepackage{graphicx}
\usepackage{caption} 
\usepackage{wrapfig}
\usepackage{enumitem}

\usepackage{xr}
\makeatletter

\newcommand*{\addFileDependency}[1]{
\typeout{(#1)}
\@addtofilelist{#1}
\IfFileExists{#1}{}{\typeout{No file #1.}}
}\makeatother

\usepackage{algpseudocode}
\usepackage{diagbox}
\usepackage{adjustbox}
\usepackage{wrapfig}
\usepackage{lipsum}
\usepackage{algpseudocode}
\usepackage{algorithm}
\usepackage{multirow}

\allowdisplaybreaks

\theoremstyle{plain}
\newtheorem{theorem}{Theorem}

\newtheorem{remark}{Remark}
\newtheorem{assumption}{Assumption}
 
\newtheorem{lemma}{Lemma}

\def \hat{\widehat}
\def \tilde{\widetilde}

\usepackage{tikz}
\usetikzlibrary{shapes,decorations,arrows,calc,arrows.meta,fit,positioning}
\tikzset{
	-Latex,auto,node distance =1 cm and 1 cm,semithick,
	state/.style ={ellipse, draw, minimum width = 0.7 cm},
	point/.style = {circle, draw, inner sep=0.04cm,fill,node contents={}},
	bidirected/.style={Latex-Latex,dashed},
	el/.style = {inner sep=2pt, align=left, sloped}
}

\theoremstyle{theoremstyle}

\newenvironment{proc1}{
  \par\medskip
  \noindent\textbf{ Stage 1.}\itshape     
}{
  \par\medskip
}

\newenvironment{proc2}{
  \par\medskip
\noindent\textbf{After Stage 1.}\itshape  ~%
}{\par\medskip}

\newenvironment{proc3}{
  \par\medskip
  \noindent\textbf{ Stage 2.}\itshape  
}{\par\medskip}

\newenvironment{proc4}{
  \par\medskip
  \noindent\textbf{After Stage 2.}\itshape  
}{\par\medskip}

\makeatletter
\newcommand{\leqnomode}{\tagsleft@true}
\newcommand{\reqnomode}{\tagsleft@false}
\makeatother

\begin{document}
	\title{ A Unified Adaptive Enrichment Design for Power Enhancement }
	\author{Junzhe Shao$^1$ \and Aibo Gong$^2$ \and Juan Shen$^3$ \and Waverly Wei$^{4}$\thanks{Correspondence: waverly@marshall.usc.edu.}}
	\date{
    $^{1}$ Division of Biostatistics, University of California, Berkeley, Berkeley, USA\\
$^{2}$ School of Economics, Peking University, Beijing, China\\
$^{3}$ Department of Statistics and Data Science, Fudan University, Shanghai, China\\
    $^4$Department of Data Sciences and Operations, University of Southern California, Los Angeles, USA\\
 \vspace{0.5cm}
}
		\maketitle

\onehalfspacing

\begin{abstract}
   Randomized controlled trials (RCTs) are the gold standard for evaluating treatment effects, but
fixed eligibility criteria and enrollment decisions can be inefficient, especially when treatment effects vary across patient subpopulations. Adaptive enrichment trials update enrollment using interim data to improve efficiency. Enrichment methods are commonly developed for two settings:
prespecified subgroups, and continuous covariates where enrollment is guided by a
learned cutoff. Many commonly used
enrichment designs adopt discontinuous rules that favor one single subgroup, which may induce ``winner’s
curse" bias if the final estimation does not properly account for the
data-dependent enrollment decision and require additional bias correction. We propose a unified framework that bridges these settings by formulating enrichment as
a regularized optimization over the enrolled covariate distribution. In a two-stage design, Stage 2
selects an enrollment mixture by maximizing a power objective while penalizing deviation from
a prespecified baseline target population through a Kullback--Leibler divergence term, providing a
smooth alternative to discontinuous pick-the-winner rules; the same formulation extends naturally to
optimizing enrollment over continuous covariates. The resulting estimand is the average treatment
effect in the trial population induced by the data-adaptive enrollment rule, so uncertainty quantification
must account for randomness in learning the optimal enrollment rule, in addition to outcome estimation.
We derive an influence-function representation for the estimated optimal enrollment rule and account for it in the final estimator, yielding an explicit asymptotic variance decomposition into decision
uncertainty and outcome-estimation uncertainty. Simulations demonstrate improved power relative to
conventional enrichment approaches while substantially reducing winner's curse bias in treatment effect
estimation.
    
    \smallskip 
     
\end{abstract}

\doublespacing

\section{Introduction}

\subsection{Motivation and contribution}
Randomized controlled trials (RCTs) are the gold standard for evaluating treatment effects, but
traditional RCTs fix the enrollment eligibility criteria and the subgroup proportions throughout the trial.
However, fixed enrollment probabilities can be statistically inefficient when treatment effects are
heterogeneous across subpopulations, potentially leading to reduced power.

Adaptive enrichment trials address this limitation by conducting the trial across multiple stages and using
interim data to update enrollment probabilities, thereby improving efficiency under treatment-effect
heterogeneity \citep{fda_enrichment_2019, frieri_design_2023, baldi_antognini_new_2023}. Enrichment methods
are commonly developed for two settings: prespecified subgroups in confirmatory trials \citep{stallard_adaptive_2014}, and continuous
covariates where enrichment is implemented by learning a cutoff (or searching over nested
subgroups) \citep{lin_inference_2021, stallard_adaptive_2023}. In practice, many enrichment designs rely on a deterministic ``pick-the-winner'' rule that enrolls
only the subgroup with the larger estimated interim effect.
Such discontinuous selection is sensitive to sampling variability when subgroup effects are close, and can
induce winner's curse bias if final estimation does not properly account for the data-dependent enrollment
decision. Moreover, adaptive enrichment alters the covariate distribution of the enrolled trial population.
When trial conclusions are intended to be interpreted relative to a prespecified baseline target population,
large population shifts based on limited trial data can complicate interpretation and motivate control of such shifts.

Motivated by these considerations, we propose a unified adaptive enrichment framework that formulates interim
enrollment as an optimization problem. In a two-stage setting, instead of deterministic subgroup selection,
the interim analysis chooses a continuous enrollment proportion by maximizing a penalized design objective.
The objective combines a power-oriented component and a Kullback--Leibler (KL) divergence penalty that
discourages deviation from a prespecified baseline subgroup prevalence.  This formulation
replaces hard winner-take-all decisions with a smooth enrollment update, and it naturally bridges the two
common enrichment settings by extending from mixtures over discrete subgroup labels to optimized enrollment
distributions over a continuous covariate space. We summarize our contributions as follows:

From a methodological perspective, we view adaptive enrichment as learning a regularized reweighting of the
enrolled covariate distribution and implement enrollment adaptation through an explicit optimization.
This replaces discrete winner-take-all selection with a continuous enrollment mixture and targets the
treatment effect in the resulting design-induced trial population, keeping the estimand aligned with the
design. The design objective couples a standardized power proxy with a KL-divergence penalty that controls
population shift. We use the KL penalty because it provides a smooth, convex, and interpretable measure of population shift from the prespecified baseline distribution, with its tuning parameter trading statistical efficiency against deviation from the baseline target population.
This design-centric regularization perspective complements,
rather than replaces, the confirmatory multiple-testing literature for prespecified subgroups: instead of
relying solely on post hoc adjustments after discrete selection, we build stability and population-shift
control into the enrollment rule itself. Furthermore, by moving from a mixture over subgroups 
to an optimized enrollment distribution over a continuous covariate space, the framework connects directly
to cutoff-based enrichment without requiring ad hoc discretization, which potentially loses information.

From a theoretical perspective, we provide guarantees that the average treatment effect (ATE) estimator
under our proposed adaptive enrichment procedure is asymptotically normal and supports valid statistical
inference (Theorem~\ref{thm:psi_formal}). A key technical challenge is that the inferential target is
design-induced, since it is defined through the optimal enrollment probability learned from interim data.
As a result, the final ATE estimator inherits uncertainty not only from estimating subgroup treatment
effects (Theorem~\ref{thm:tau_j_formal}), but also from estimating the optimal enrollment probability itself
(Lemma~\ref{lemma:p_star_formal}). We derive an influence-function representation for the data-adaptive
optimal enrollment probability and incorporate it into the final estimator, yielding an asymptotic variance
expression that decomposes uncertainty into design-decision uncertainty from learning the optimal enrollment
probability and outcome-estimation uncertainty from estimating subgroup treatment effects. From our simulation studies (Section \ref{sec:sim}), we report scenario-specific operating characteristics for the proposed and benchmark
designs. Overall, the proposed approach yields an interpretable adaptive enrichment framework with valid inference under the stated regularity conditions.

\subsection{Literature review}
Enrichment designs have a long history in clinical trials. In practice, enrichment typically follows a pick-the-winner logic: Stage 1 data are used to identify the most promising population or subpopulation, after which enrollment or analysis is adaptively modified in later stages to concentrate on patients most likely to benefit, while maintaining confirmatory error control.
A key distinction is whether subgroups are predefined. When they are, enrichment can be formulated as a multiple testing problem across overall and prespecified subgroups coupled with adaptive enrollment. In this setting, regulatory guidance emphasizes pre-specification, accounting for adaptation-induced variability, and strong familywise error rate (FWER) control \citep{fda_enrichment_2019}, with closed testing procedures serving as a canonical approach \citep{marcus_closed_1976}.

In the predefined-subgroup literature, \citet{rosenblum_optimizing_2011} propose a general framework for two-stage enrichment RCTs with interim enrollment adaptation, achieving asymptotic strong FWER control and power gains for benefiting subgroups without additional selection penalties, but requiring multivariate normal probability calculations. \citet{stallard_adaptive_2014} provide a broader regulatory-focused perspective on confirmatory adaptive trials that identify and enrich promising predefined subgroups. \citet{rosenblum_multiple_2016} develop group-sequential multiple testing procedures that leverage covariance and alpha reallocation to improve power while strongly controlling FWER. Building upon this line of work, 
\citet{rosenblum_optimal_2020} jointly optimize interim enrollment rules and final multiple testing procedures, using sparse linear programming to minimize expected sample size under power and FWER constraints.

A second line of work considers continuous covariates, where enrichment involves searching over many nested, data-adaptive subgroups. \citet{simon_adaptive_2013,simon_inference_2017} develop designs that adapt eligibility criteria to enrich for likely responders, typically targeting a strong null to facilitate error control under extensive searching but can be conservative. \citet{lin_inference_2021} propose a two-stage, threshold-based parametric enrichment strategy for continuous biomarkers, using random norming to enable less conservative inference after adaptive screening. 
 \citet{magnusson_group_2013} study group-sequential enrichment with subgroup identification, emphasizing post-selection estimation and likelihood-based efficiency within a strong FWER-controlled framework. 
\citet{stallard_adaptive_2023} develop valid testing for two-stage adaptive enrichment with nested subgroups, exploiting an equivalence to group-sequential statistics and closed testing to achieve strong FWER control under a monotone-effect assumption.

Bias correction is central when subgroup selection is coupled with effect estimation. \citet{simon_inference_2017} show that prespecifying the selection algorithm enables resampling-based corrections, such as bootstrap, to mitigate winner’s curse bias in subgroup effect estimation. Similarly, \citet{zhang_treatment_2018} demonstrate that prespecified selection rules allow cross-validation and bootstrap methods to correct selection bias and support valid post-selection inference. 
These approaches primarily intervene at the estimation stage while retaining the underlying hard subgroup-selection or pick-the-winner (PTW) enrollment rule. Therefore, they can reduce bias in the estimated effect for the selected target, but they do not alter the subsequent stage's enrichment decision or recover information from subgroups excluded after interim selection. Our approach regularizes the enrichment decision itself by replacing the discontinuous PTW rule with a continuous, variance-aware enrollment mixture. Thus, bias-corrected PTW procedures and the proposed method address complementary components of the adaptive procedure.

While enrichment designs are largely power-motivated, the choice of objective function and optimization strategy can substantially affect the adaptation rule. More broadly, statistical efficiency can also be improved by leveraging additional sources of information, such as historical data in other evaluation settings \citep{ruan2026hero}. Common objectives include selecting the subgroup with the largest Wald statistic or estimated treatment effect, maximizing impact criteria such as effect size times prevalence, or maximizing standardized evidence of effect modification, as reviewed in \citet{stallard_adaptive_2023}. These objectives implicitly trade off confirmatory power, subgroup identification accuracy, and the size and interpretability of the target population.

\section{Problem setup for classical enrichment design} \label{sec:method}

To clarify the problem setup, we begin with a two-stage adaptive enrichment design involving a binary covariate and a binary treatment. For each stage $t\in\{1,2\}$, there are $N^{(t)}$ subjects enrolled. The total sample size is $N = N^{(1)} + N^{(2)}$. For each enrolled subject $i\in\{1,2,\ldots,N\}$, we record their treatment status $D_i \in \{0,1\}$, enrollment stage $S_i\in\{1,2\}$, covariate $X_i\in\{1,2\}$, and observed outcome $Y_i$.  Under the potential outcome framework \citep{splawa-neyman_application_1923,rubin_estimating_1974}, let $Y_i(d)$ denote the potential outcome of subject $i$ under treatment $d\in\{0,1\}$. Aligned with the existing literature on experimental design, we operate in a setting where the potential outcomes are defined as $Y_1(d), \ldots, Y_N(d)$. Enrichment designs typically maintain constant treatment assignment probabilities across the two stages. Formally, this means that $
    \mathbb{P}(D_i = d \mid X_i = j, S_i = t) = \mathbb{P}(D_i = d \mid X_i = j),  \text{for } i = 1, \ldots, N,$
where $j\in\{1,2\}$.
However, recruitment criteria may shift between stages, leading to different probabilities of recruiting subjects with the same covariate: 
\begin{align}
    \mathbb{P}(X_i = j\mid S_i = 1 ) \neq \mathbb{P}(X_i = j\mid S_i = 2), \quad  \text{for } i = 1, \ldots, N.
\end{align}
To prepare for further discussion, we define 
\begin{align}
   p^{(s)}_1 := \mathbb{P}(X_i = 1 \mid S_i = s  ), \ p_2^{(s)}:= \mathbb{P}(X_i = 2 \mid S_i = s) ,\
 p^{(s)}_1 + p^{(s)}_2 = 1,
\end{align}
where $s\in \{1,2\}$.
Lastly, the primary parameter of interest in an enrichment design is typically the average treatment effect in the ``winning" subgroup (without loss of generality, we assume larger effects indicate better treatments):
\begin{align}
    \tau_{\max} &:= \max\{\tau_1, \tau_2\}, \\
    \tau_j &:= \mathbb{E}\left[Y_i(1) - Y_i(0) \mid X_i = j,S_i = t \right] \notag
= \mathbb{E}\left[Y_i(1) - Y_i(0) \mid X_i = j\right].
\end{align}
A secondary parameter of interest may be the overall average treatment effect (ATE) in the entire trial. Let $p_j = \mathbb{P}(X_i=j|S_i\leq 2)$ be the cumulative enrichment proportion over two stages for subgroup $j$, $j=1,2$. Let $\kappa = N^{(1)}/N$, the planned cumulative mixture is $p_j = \kappa p_j^{(1)}+(1-\kappa)p_j^{(2)}$, and the realized sample proportion is its empirical analogue. The ATE in the trial can be defined as $ \tau(p_1,p_2) = \tau_1 p_1 + \tau_2 p_2$. Because we only consider a two-subgroup setting such that $p_2=1-p_1$, we can further simplify the above ATE as
$\tau(p_1) = \tau_1 p_1 + \tau_2 (1-p_1).$

The classical enrichment design usually aims to determine the winning subgroup and subsequently infer its ATE. Investigators commonly conduct an interim analysis between Stage 1 and Stage 2, eliminating the subgroup exhibiting the more negligible estimated effect. In a simplified setting, the enrollment decision in the second stage is made based on the first stage data; for example,  
\begin{align}
\label{eq::prob}
     \hat{p}^{(2)}_1 = \mathbf{1} ( \hat\tau_1^{(1)} >  \hat\tau_2^{(1)} ), 
\end{align}
where $ \hat\tau_1^{(s)} $ denotes estimate of $\tau_1$ using the data collected in the stage $s$.

\begin{remark}[Issues within classical enrichment designs]\label{remark:issues within classical enrichment designs}
  In classic enrichment designs, the enrollment decision is based on comparing the Stage 1 estimates $\hat\tau_1^{(1)}$ and $\hat\tau_2^{(1)}$. One proceeds by enrolling only subgroup 1 if $\hat\tau_1^{(1)} > \hat\tau_2^{(1)}$, and analogously for subgroup 2. The corresponding oracle selection rule is
$p^{(2)}_1 = \mathbf{1} ( \tau_1 >  \tau_2 ).$
In practice, however, decisions are based on the empirical counterpart in Eq (\ref{eq::prob}). Since the estimates $\hat{\tau}_j^{(1)}$ are subject to sampling variability, the estimated enrollment probability $\hat{p}_1^{(2)}$ may deviate from the oracle enrollment probability $p_1^{(2)}$. 
For example, even when $\tau_1 > \tau_2$ (implying $p_1^{(2)}=1$), random error may lead to $\hat{\tau}_1^{(1)} < \hat{\tau}_2^{(1)}$, yielding the incorrect decision $\hat{p}_1^{(2)}=0$. The problem is particularly severe when $\tau_1$ lies in a $\sqrt{N}$-neighborhood of $\tau_2$. In such settings, the adaptive enrollment decision may select the suboptimal subgroup in Stage 2, so $\hat{p}_1^{(2)}$ fails to converge to $p_1^{(2)}$. A related issue arises when subgroup 1 is selected because $\hat{\tau}_1^{(1)}$ is upwardly biased by positive sampling error. Incorporating the same Stage 1 data into the final estimation of $\tau_1$ induces a systematic overestimation of the true effect, a phenomenon often referred to as the winner’s curse bias, which has been well recognized in adaptive enrichment literature \citep{simon_inference_2017,zhang_treatment_2018}. 
\end{remark}

\section{Proposed enrichment design} \label{sec:proposed-design}

\subsection{Oracle enrichment design}

In this section, we start by introducing the oracle enrichment design. By the ``oracle", we refer to the setting where we have perfect knowledge of the true underlying parameters. The oracle problem aims to identify the optimal proportion of subgroup 1 across two stages, denoted $p_1^*$, that maximizes a penalized objective function balancing statistical efficiency against distributional shift. We write $q_1:= p_1^{(1)}$ for the prespecified
Stage 1 baseline prevalence. Let \(\kappa:=N^{(1)}/N\). Since Stage~1 enrolls with fixed prevalence \(p_1^{(1)}\) and Stage~2 must satisfy
\(p_1^{(2)}\in[0,1]\), the two-stage mixture
\(
p_1=\kappa p_1^{(1)}+(1-\kappa)p_1^{(2)}
\)
is restricted to this feasible interval $
    \mathcal P_\kappa
=\Big[\kappa p_1^{(1)},\ 1-\kappa(1-p_1^{(1)})\Big]\subset(0,1)$.
We therefore solve the oracle problem over \(p_1\in \mathcal P_\kappa\).
\begin{align}\label{eq:oracle_objective_discrete}
    p_1^* = \arg \max_{p_1 \in \mathcal P_\kappa} &\Biggl\{\underbrace{\frac{p_1\,\tau_1 + (1-p_1)\,\tau_2}{\sqrt{\,p_1\,V_1 + (1-p_1)\,V_2}}}_{(\text{i})}\\
    - &\underbrace{\lambda\, \left[p_1 \log\left(\frac{p_1}{p_1^{(1)}}\right)  + (1-p_1) \log\left(\frac{1-p_1}{1-p_1^{(1)}}\right)\right]}_{(\text{ii})}\Biggr\}, \nonumber
\end{align}
where   $$V_j := \frac{\sigma_j^2(1)}{e_j} + \frac{\sigma_j^2(0)}{1-e_j}, \quad \sigma_j^2(d):= \mathbb{V}[Y_i(d) \mid X_i = j],$$
and $d\in\{0,1\}, j\in\{1,2\}$, and $e_j := \mathbb{P}(D_i=1|X_i=j)$.
Here, $\lambda \ge 0$ is a penalty parameter that controls the strength of regularization. $V_j$ characterizes the asymptotic variance of the treatment effect estimator within subgroup $j$.

Term (i) represents the standardized ATE used in the enrichment optimization. Its numerator represents the overall ATE, while its denominator is the asymptotic standard deviation of the corresponding estimator under the subgroup proportion $p_1$. Intuitively, when $\tau_1 > \tau_2$, increasing $p_1$ increases the numerator, but the denominator penalizes oversampling high-variance subgroups: If subgroup 1 has a larger variance, increasing $p_1$ inflates variance and can reduce the term (i).

Term (ii) is a Kullback-Leibler (KL) penalty that regularizes the subgroup proportion $p_1$ towards its prevalence in the general population observed in Stage 1.  The KL penalty provides a smooth, convex, and interpretable measure of deviation from a prespecified baseline population, allowing its tuning parameter to balance statistical efficiency against the extent of adaptive enrichment. This regularization stabilizes Stage~2 enrollment decisions by reducing sensitivity to noisy Stage~1 estimates.
The penalty is zero at 
$p_1 = p_1^{(1)}$, which is locally quadratic and strictly convex. The tuning parameter $\lambda$ controls the trade-off between the design objective (term (i)) and robustness to distributional shift (term (ii)). Furthermore, $\lambda = 0$ yields a purely utility maximization objective, while large $\lambda$ pushes $p_1^*$
towards $p_1^{(1)}$.

Our oracle formulation addresses the issues in the classical enrichment designs in Remark \ref{remark:issues within classical enrichment designs} by replacing the discontinuous ``pick-the-winner" rule with a continuous allocation $p_1^*$, obtained from a regularized objective. Rather than collapsing Stage 2 enrollment to $\{0,1\}$, which is highly sensitive to Stage 1 estimation noise, we choose $p_1\in[0,1]$ to maximize a standardized effect (term (i)) that explicitly accounts for subgroup-specific variability through $V_j$, mitigating over-enrichment into high-variance subgroups that would reduce power. The KL penalty further pushes $p_1$ toward the population prevalence $p_1^{(1)}$. Together, these features stabilize the Stage~2 decision when $\tau_1-\tau_2 = O(N^{-1/2})$, reducing the risk of locking into a suboptimal subgroup purely due to sampling variability.

\subsection{Proposed two-stage adaptive enrichment design}

In this section, we move from the oracle formulation to the practical implementation of our adaptive enrichment design. We illustrate the proposed methodology in a two-stage setting. We introduce an additional notation by denoting the treatment assignment probability for subgroup $j$ at Stage $t$ as $e_j^{(t)}$.

\begin{proc1}
    Enroll $N^{(1)}$ subjects in Stage 1 with enrollment probability $p_j^{(1)}$, and randomize each subgroup to treatment with probabilities $e_j^{(1)} = 0.5$, $j=1,2$.
\end{proc1}
In Stage 1, subjects are enrolled according to the prespecified baseline prevalence $q_j = p_j^{(1)}$. Stage 1 treatment assignment probability is defined as $e_j^{(1)} := \mathbb{P}(D_i=1|X_i=j, S_i = 1)$ and is fixed  at 0.5.

\begin{proc2}
Using Stage 1 data, estimate the optimal enrollment probability $\hat{p}_1^*$ by solving the empirical version of the oracle optimization problem as 
\begin{align}\label{eq:empirical objective function}
    \hat{p}_1^* &= \arg\max_{p_1\in \mathcal P_\kappa} \Biggl\{\frac{p_1\,\hat{\tau}_1^{(1)} + (1-p_1)\,\hat{\tau}_2^{(1)}}{\sqrt{\,p_1\,\hat{V}_1^{(1)} + (1-p_1)\,\hat{V}_2^{(1)}}} 
    \notag\\
    &- \lambda\, \left[p_1 \log\left(\frac{p_1}{p_1^{(1)}}\right) + (1-p_1) \log\left(\frac{1-p_1}{1-p_1^{(1)}}\right)\right]\Biggr\},
\end{align}
where $\lambda$ is the tuning parameter prespecified for the trial,  
\begin{align}\label{eq:tau-hat-stage-1-estimate}
    \hat{\tau}_j^{(1)} &= {  \hat{\mu}_j^{(1)}(1)  -  \hat{\mu}_j^{(1)}(0) },\quad \hat{V}_j^{(1)} = \frac{\hat{\sigma}_j^2(1)^{(1)}}{e_j^{(1)}} + \frac{\hat{\sigma}_j^2(0)^{(1)}}{1-e_j^{(1)}},
\end{align}
and
\begin{align}\label{eq:hat-mu-j and hat-sigma-j}
    \hat{\mu}_j^{(1)}(d) 
    &= \frac{\sum_{i: S_i = 1} \mathbf{1}(X_i = j)\,\mathbf{1}(D_i = d)\,Y_i}
            {\sum_{i: S_i = 1} \mathbf{1}(X_i = j)\,\mathbf{1}(D_i = d)}, \\ \hat{\sigma}_j^2(d)^{(1)}
    &= \frac{\sum_{i: S_i = 1} \mathbf{1}(X_i = j)\,\mathbf{1}(D_i = d)\,
            \bigl(Y_i -{ \hat{\mu}_j^{(1)}(d)}\bigr)^2}
            {\sum_{i: S_i = 1} \mathbf{1}(X_i = j)\,\mathbf{1}(D_i = d) - 1},\nonumber
\end{align}
\end{proc2}

\begin{proc3}
Enroll subgroup 1 with probability $\hat{p}_1^{(2)}$, where
   $\hat{p}_1^{(2)} = \frac{\hat{p}_1^* - \kappa p_1^{(1)}}{1-\kappa}, \kappa = \frac{N^{(1)}}{N}$, enroll $N^{(2)}$ subjects with subgroup enrollment probabilities $\hat{p}_j^{(2)}$, and assign treatment with probability $e_j^{(2)} := \mathbb{P}(D_i=1|X_i=j, S_i=2)= \frac{1}{2}$. 
\end{proc3}
Because $\hat{p}_1^*\in\mathcal{P}_{\kappa}$, the plug-in Stage 2 probability $\hat{p}_1^{(2)}$ lies in $[0,1]$. The
realized final enrollment proportion need not equal $\hat{p}_1^*$ exactly. Instead, it is the
empirical analogue of the planned cumulative mixture.

\begin{proc4}
Estimate the ATE using data collected from the two stages as
\begin{align}\label{eq:proposed ate estimator}
     \hat{\tau} = \hat{p}_1^* \hat{\tau}_1 + (1-\hat{p}_1^*) \hat{\tau}_2,
\end{align}
where
    $\hat{\tau}_j = {\hat{\mu}_j{(1)}- \hat{\mu}_j{(0)}}.$
\end{proc4}
After Stage 2, we estimate the ATE using the data from both stages. 
The estimator $\hat{\tau}$ targets $\tau(p_1^*) = p_1^* \tau_1 + (1-p_1^*) \tau_2$, with inference following directly from Theorem~\ref{thm:psi_formal} in Section~\ref{sec:theory}.
Using all collected data (\(S_i\le2\)), we define the arm-specific means and variances in subgroup \(j\) by
\begin{align}
\hat\mu_j(d)
&:=\frac{\sum_{i:S_i\le2}\mathbf{1}(X_i=j)\mathbf{1}(D_i=d)\,Y_i}
{\sum_{i:S_i\le2}\mathbf{1}(X_i=j)\mathbf{1}(D_i=d)},\\
\hat\sigma_j^2(d)
&:=\frac{\sum_{i:S_i\le2}\mathbf{1}(X_i=j)\mathbf{1}(D_i=d)\,\bigl(Y_i-\hat\mu_j(d)\bigr)^2}
{\sum_{i:S_i\le2}\mathbf{1}(X_i=j)\mathbf{1}(D_i=d)-1},\ d\in\{0,1\}. \nonumber
\end{align}

For brevity, we write
$O_i=(Y_i,D_i,X_i,S_i)$ for the observed data on subject $i$. The influence-function based variance estimator is
\begin{align}\label{eq:var_if_hat_tau}
\widehat{\mathbb{V}}(\hat\tau)
:=
\frac{1}{N(N-1)}\sum_{i=1}^N\Bigl(\hat\psi_{\tau}(O_i)-\bar\psi_\tau\Bigr)^2,
\qquad
\bar\psi_\tau:=\frac{1}{N}\sum_{i=1}^N \hat\psi_{\tau}(O_i),
\end{align}
where
\begin{align}\label{eq:psi_tau_hat}
\hat\psi_{\tau}(O_i)
&:=
(\hat\tau_1-\hat\tau_2)\hat\psi_{p}(O_i)
+
\hat p_1^*\,\hat\psi_{1}(O_i)
+
(1-\hat p_1^*)\,\hat\psi_{2}(O_i),\\
\hat\psi_j(O_i)&=\frac{N\,\mathbf{1}(X_i=j)D_i}{\sum_{i=1}^N\mathbf{1}(X_i=j)D_i}\,\bigl(Y_i-\hat\mu_j(1)\bigr) \notag\\
&- \frac{N\,\mathbf{1}(X_i=j)(1-D_i)}{\sum_{i=1}^N\mathbf{1}(X_i=j)(1-D_i)}\,\bigl(Y_i-\hat\mu_j(0)\bigr), 
\end{align}
and 
\begin{align}\label{eq:psi_p_stage1_impl}
\hat\psi_{p}(O_i)
=
-{\frac{\mathbf{1}(S_i=1)}{\kappa}}
\Bigl\{\partial_{pp} M_{N^{(1)}}(\hat p_1^*)\Bigr\}^{-1}
\Bigl[\nabla_\theta\,\partial_p F(\hat\theta^{(1)},\hat p_1^*)\Bigr]^\top
\hat\Phi_i^{(1)},
\end{align}
where $M_{N^{(1)}}(\hat{p}_1^*)$ is the Stage~1 empirical objective in \eqref{eq:empirical objective function},
$$\hat\theta^{(1)}=(\hat\tau^{(1)}_1,\hat\tau^{(1)}_2,\hat V^{(1)}_1,\hat V^{(1)}_2)^\top, \ F(\hat{\theta}^{(1)}, \hat{p}_1^*):= \frac{\hat{p}^*_1\,\hat{\tau}^{(1)}_1 + (1-\hat{p}^*_1)\,\hat{\tau}^{(1)}_2}{\sqrt{\,\hat{p}^*_1\,\hat{V}^{(1)}_1 + (1-\hat{p}^*_1)\,\hat{V}^{(1)}_2}}$$
and $\hat\Phi_i^{(1)}$ is the empirical influence-function vector for $\hat\theta^{(1)}$. The detailed calculations are included in Supplementary Materials.
A two-sided $(1-\alpha)$ Wald confidence interval is then $$
    \hat\tau \ \pm\ z_{1-\alpha/2}\,\sqrt{\widehat{\mathbb{V}}(\hat\tau)}.$$

\section{Generalization of the proposed design} \label{sec:generalization-j}

In this section, we extend the proposed adaptive enrichment design along two directions:
(i) jointly adapting both enrollment probabilities and treatment assignment probability and 
 (ii) incorporating continuous covariates, while generalizing the two-subgroup case to multiple subgroups.

\subsection{Simultaneous optimization of treatment assignments}\label{subsec:simultaneous optimization}

In the oracle problem formulation of Section \ref{sec:proposed-design}, the within-subgroup treatment assignment probability $e_j$  was fixed at $0.5$, even though the variance contribution depends on $e_j$. To leverage this additional degree of freedom and improve efficiency, we extend the framework to jointly optimize $e_j$ and the enrollment probability $p_j$.

We consider a multi-subgroup setting with subgroups $j=1,\ldots, J$. Let $\mathbf{p} := (p_1, \dots, p_J)^\top$ denote enrollment probabilities and 
$\mathbf{e} = (e_1, \dots, e_J)^\top$ denote treatment assignment probabilities, where $e_j:= \mathbb{P}(D_i=1|X_i=j)\in(0,1)$. The asymptotic variance of the subgroup treatment effect depends on $e_j$, given by $
    V_j(e_j) = \frac{\sigma_{j}^2(1)}{e_j} + \frac{\sigma_{j}^2(0)}{1-e_j}.$
We now formulate the oracle problem to obtain the optimal pair $(\mathbf{p}^*, \mathbf{e}^*)$. Let $d_B(e||e_0) = e\log(e/e_0)+(1-e)\log\{(1-e)/(1-e_0)\}$ denote the binary KL divergence. The joint objective is
\begin{align} \label{eq:joint_obj}
    (\mathbf{p}^*, \mathbf{e}^*) &= \mathop{\arg \max}_{\mathbf{p}, \mathbf{e}} \Bigg\{ \underbrace{\frac{\sum_{j=1}^J p_j \tau_j}{\sqrt{\sum_{j=1}^J p_j V_j(e_j)}}}_{\text{(i)}} - \underbrace{\lambda_1 \sum_{j=1}^J p_j \log\left(\frac{p_j}{p_j^{(1)}}\right) - {\lambda_2 \sum_{j=1}^J (p_j-\kappa p_j^{(1)}) d_B(e_j||e_j^{(1)})}}_{\text{(ii)}}  \Bigg\}.
\end{align}
 The tuning parameters $\lambda_1$ and $\lambda_2$ govern the trade-off between power maximization and preservation of the general population mixture and randomization, respectively.

Term (i) in \eqref{eq:joint_obj} generalizes the enrichment objective to multiple subgroups. The numerator is the ATE under enrollment $\mathbf{p}$, and the denominator retains the Stage 1 and Stage 2 variance contributions. Increasing weight on subgroups with larger  $\tau_j$ raises the numerator but also amplifies their variance contribution. In contrast, $e_j$ affects $V_j(e_j)$ without changing $\tau_j$. Jointly optimizing the pair can therefore enhance power, particularly under subgroup-specific heteroskedasticity.

Term (ii) consists of two penalties. The $\lambda_1$ penalty generalizes the two-subgroup case to balance distributional shift, shrinking the trial population toward the baseline prevalence $(p_1^{(1)},\ldots, p_J^{(1)})$. The $\lambda_2$ penalty  uses the binary KL divergence to limit departure from $e_j^{(1)}$, weighted by the Stage 2 subgroup contribution. When $e_j= e_j^{(1)}$ or $\lambda_2 =0$, the randomization penalty vanishes.

We now describe the practical implementation, omitting Stage~1 details since they are analogous to the two-subgroup case. We adopt a two-step optimization for simplicity.

\begin{proc2}
    Using  Stage 1 data, we estimate $\hat{\tau}_j^{(1)}$ and $\hat{V}_j^{(1)}$ for $j=1,\ldots, J$ as in Eq \eqref{eq:tau-hat-stage-1-estimate}. We then set the Stage 2 randomization probability to the Neyman allocation ratio \citep{neyman_two_1934}:
        $\hat{e}_j^{*} = \frac{\hat{\sigma}_{j}^{(1)}(1)}{\hat{\sigma}_{j}^{(1)}(1) + \hat{\sigma}_{j}^{(1)}(0)}$,  compute $\hat{V}_j(\hat{e}_j^{*})$ as  $\hat{V}_j(\hat{e}^{*}_j) = \frac{\hat{\sigma}_j^2(1)^{(1)}}{\hat{e}_j^{*}} + \frac{\hat{\sigma}_j^2(0)^{(1)}}{1-\hat{e}_j^{*}}.$ Plugging $\hat{V}_j(\hat{e}_j^{*})$ into the objective function in Eq \eqref{eq:joint_obj}
 to solve for the optimal enrollment probabilities $\hat{\mathbf{p}}^*$:
\begin{align}
    \hat{\mathbf{p}}^* &= \mathop{\arg \max}_{\mathbf{p}} \Biggl\{ \frac{\sum_{j=1}^J p_j \hat{\tau}_j^{(1)}}{\sqrt{\sum_{j=1}^J p_j \hat{V}_j(\hat{e}_j^*)}}- \lambda_{1} \sum_{j=1}^J p_j \log\left(\frac{p_j}{p_j^{(1)}}\right) \Biggr\}.
\end{align}
\end{proc2}

Because directly solving the empirical joint optimization in \eqref{eq:joint_obj} can be computationally intensive, we adopt a decoupled strategy. We first optimize
$\mathbf{e}$ by minimizing $V_j(e_j)$ within each subgroup, yielding the Neyman allocation and efficient within-subgroup variance reduction. We then optimize $\mathbf{p}$ given these variances. This stepwise rule is exact when $\lambda_2 = 0$ and the planned weighted effect is positive. For $\lambda_2>0$, the penalty couples $\mathbf{p}$ and $\mathbf{e}$. We therefore use the
$\lambda_2=0$ stepwise solution only as an initialization for
multiple-start constrained optimization.

\begin{proc3}
    In Stage 2, subgroup $j$ is enrolled with probability $\hat{p}_j^{(2)}$ and assigned treatment with probability $\hat{e}_j^{(2)}$, where 
$
      \hat{p}_j^{(2)} = \frac{\hat{p}_j^* - \kappa p_j^{(1)}}{1-\kappa}, \ \hat{e}_j^{(2)} = \hat{e}_j^*$
    $ \text{for } j = 1, \dots, J$, with $\kappa = N^{(1)}/N$. 
\end{proc3}

\begin{proc4} After Stage 2, the ATE is estimated as $
     \hat{\tau} = \sum_{j=1}^J \hat{p}_j^* \hat{\tau}_j$,
where $$ \hat{\tau}_j = \frac{\sum_i \mathbf{1}(X_i=j)D_iY_i/e_j^{(S_i)}}{\sum_i \mathbf{1}(X_i=j)D_i/e_j^{(S_i)}}- \frac{\sum_i \mathbf{1}(X_i=j)(1-D_i)Y_i/(1-e_j^{(S_i)})}{\sum_i \mathbf{1}(X_i=j)(1-D_i)/(1-e_j^{(S_i)})}.$$
\end{proc4}

Similar to the two-subgroup setting, the variance estimator has the same form as in Eq \eqref{eq:var_if_hat_tau}. The only difference lies in the estimator $\hat\psi_\tau(O_i)$, where
\begin{align}\label{eq:psi_tau_decomp_multi}
\hat\psi_\tau(O_i)
=
\sum_{j=1}^{J-1}(\hat\tau_j-\hat\tau_J)\,\hat\psi_{p,j}(O_i)
\;+\;
\sum_{j=1}^J \hat p_j^*\,\hat\psi_j(O_i).
\end{align}
Here $\hat\psi_{p,j}(O_i)$ is the $j$th component of empirical influence function for the Stage~1 optimizer, and $\hat\psi_j(O_i)$ is the empirical influence function for the subgroup ATE estimator
$\hat\tau_j$. A two-sided $(1-\alpha)$ Wald confidence interval is then $
    \hat\tau \ \pm\ z_{1-\alpha/2}\,\sqrt{\widehat{\mathbb{V}}(\hat\tau)}$.

Although the treatment assignment probabilities may vary across stages,
treatment remains randomized because the stage- and subgroup-specific
probabilities \(e_j^{(t)}\) are known and bounded away from zero and one.
However, varying \(e_j^{(t)}\) makes treatment assignment associated with
stage in the pooled sample. Therefore, unmodeled outcome drift
may induce bias if stage is ignored. The present theory therefore relies on
the assumption that the conditional outcome moments remain
stable across stages within each subgroup. If outcome drift is plausible, a
stage-adjusted analysis using the known stage-specific randomization
probabilities should be prespecified.

\subsection{Continuous covariates setting}

For the designs proposed in the previous sections, subgroups were defined by a discrete covariate 
$X$. We now extend the framework to allow
$X$ to be continuous, with $X\in\mathcal{X}$, without predefined subgroup partitions.

We formulate the oracle problem for continuous covariates. Let $Q$ denote the covariate distribution in the general population, and let $g(X)$ be a user-specified measurable function representing the contribution of enrolling a subject with covariates 
$X=x$. The choice of $g$ reflects the trial objective; for example, 
$g(x)= \frac{\tau(x)}{ \sqrt{V(x)}}$ is the utility used to rank covariate values for enrollment, where $\tau(x)$ and $V(x)$ denote the treatment effect and the variance of covariate strata $x$ respectively. 
or $g(x)=\tau(x)$ to target regions with large treatment effects.  We now formulate the oracle problem as
\begin{align}\label{eq:continuous covariate oracle problem}
P^{(2),*} \in \arg\max_{P\ll Q}\Bigl\{\underbrace{\mathbb{E}_P[g(X)]}_{\text{(i)}} - \underbrace{\lambda D_{KL}(P \Vert Q)}_{\text{(ii)}}\Bigr\},
\end{align}
where $\lambda>0$ controls the strength of regularization and $
D_{\mathrm{KL}}(P\|Q)=\int \log\!\left(\frac{dP}{dQ}\right)\,dP$ is the Kullback-Leibler divergence. Let $P^{(2),*}$ denote the optimal Stage 2 enrollment distribution solving the oracle optimization problem. The final two-stage trial distribution is $P^* = \kappa Q +(1-\kappa) P^{(2),*}$. The target parameter in the continuous-covariate setting is: 
$\tau(P^*):=\mathbb{E}_{P^*}[\tau(X)] = \int \tau(x) dP^*(x)$.

Term (i), $\mathbb E_P[g(X)]$ represents the average utility per enrolled subject under enrollment distribution $P$. By upweighting covariate regions with large $g(x)$, the design increases this expectation and thus selects subjects expected to contribute most to the trial objective. In particular, when  
 $g(x)=\tau(x)/\sqrt{V(x)}$, maximizing $\mathbb E_P[g(X)]$ prioritizes covariate regions with higher standardized power.

Term (ii) penalizes deviation of the enrolled distribution $P$ from the general population distribution $Q$. The KL penalty preserves the support of the baseline distribution and produces an explicit exponential-tilting enrollment rule. It enforces support constraints such that enrichment can only reweight covariates present under $Q$, and discourage large shifts when the signal is weak, and controls population shift. With a clear signal and a small penalty, concentrated enrichment can be appropriate. The tuning parameter controls this trade-off, that is, as $\lambda$ goes to 0, $P^{(2),*}$ concentrates on regions with large $g(x)$. As $\lambda$ increases, $P^{(2),*}$ approaches $Q$,  and hence $P^*$ also approaches $Q$.

Furthermore, the solution $P^{(2),*}$ to the oracle optimization problem is an exponential tilting of $Q$\citep{csiszar_i-divergence_1975}, given by the Donsker–Varadhan variational formula:
\begin{align}\label{eq:continuous covariate oracle problem solution}
    \frac{dP^{(2),*}}{dQ}(x) = \frac{e^{g(x)/\lambda}}{\mathbb{E}_Q\left[e^{g(X)/\lambda}\right]},
\end{align}
which defines the optimal enrollment distribution. This corresponds to a soft enrichment rule, where covariate values with larger $g(x)$ are multiplicatively upweighted relative to $Q$ by $e^{g(x)/\lambda}$. The resulting policy is the continuous analogue of a softmax rule, closely related to distributional optimization and reinforcement learning frameworks \citep{montesuma_recent_2024}. A key point is that $P^*$ depends on the choice of $g$. The closed-form solution in \eqref{eq:continuous covariate oracle problem solution} applies because the objective enters linearly as $\mathbb E_P[g(X)]$; thus its interpretation and guarantees rely on $g(\cdot)$ correctly encoding the trial objective. For example, in the discrete case  $\mathbb{E}_P[g(X)] = p_1 g(1) + (1-p_1) g(0)$, the objective from the previous section cannot be directly accommodated in Eq \eqref{eq:continuous covariate oracle problem solution} unless the variances in two subgroups are equal ($V_1=V_2$) or we focus only on subgroup effect.

In what follows, we shall illustrate the implementation procedures under the continuous covariate setting:

\begin{proc1}
     Enroll $N^{(1)}$ subjects by sampling $X_i \sim Q$, $i=1,\ldots, N^{(1)}$, and assign treatment with probability $\frac{1}{2}$.
\end{proc1}

\begin{proc2}
    Estimate $g(x)$ from Stage 1 data to obtain $\hat{g}^{(1)}(x)$, then solve the empirical optimization problem using the Stage 1 empirical distribution $Q_{N_{(1)}}$:
$$\hat{P}^{(2)} \in \mathop{\arg\max}_{P\ll Q_{N^{(1)}}} \left\{ \mathbb{E}_P[\hat{g}^{(1)}(X)] - \lambda D_{KL}(P \Vert Q_{N^{(1)}}) \right\}.$$
\end{proc2}
This step typically uses nonparametric methods to estimate components of $g(x)$, such as the CATE $\tau(x)$ and conditional variances $\sigma^2(d,x)$. 
Solving the empirical problem yields the Stage~2 target distribution via the density ratio  $\frac{d\hat{P}^{(2)}}{dQ_{N^{(1)}}}(x) \propto \exp({\hat{g}^{(1)}(x)/{\lambda}})$, implemented by rejection sampling or inverse-probability tilting. As in the discrete case, we then calibrate $\hat{P}^* = \kappa Q + (1-\kappa)\hat{P}^{(2)}$.

\begin{proc3}
 Enroll $N^{(2)}$ subjects by sampling from $\hat{P}^{(2)}$, and assign treatment with probability $\frac{1}{2}$. 
\end{proc3}

\begin{proc4}
 Keep the Stage 1-induced distribution  $\hat{P}^*$ fixed, estimate $\tau(X)$ using the full
trial data with methods that account for adaptive sampling, and then compute the ATE as
    $\hat{\tau}(\hat{P}^*) = \int \hat{\tau}(x) d\hat{P}^*(x).$ 
\end{proc4}

The integral can be evaluated via Monte Carlo or the empirical mixture weights
$\frac{d\hat{P}^*}{dQ_{N_{(1)}}}$. Estimating $g(x)$ nonparametrically using the full dataset requires methods that account for the adaptive sampling induced by $\hat{P}^*$. A detailed operational acceptance–
rejection procedure is given in Algorithm  1 of the Supplementary Materials.

\section{Theoretical investigation}
\label{sec:theory}

In this section, we study the theoretical properties of the proposed design. Lemma~\ref{lemma:p_star_formal}
establishes consistency and 
a $\sqrt{N^{(1)}}$-rate asymptotic linear representation for the estimated optimal enrichment proportion. Theorem~\ref{thm:tau_j_formal} shows that subgroup treatment effect estimators remain asymptotically normal despite adaptive enrollment, and Theorem~\ref{thm:psi_formal} combines these results via the
functional delta method to obtain asymptotic normality of the overall ATE estimator for the optimal trial population.

We summarize the key takeaways as follows: First, the estimated enrollment probability is asymptotically linear, enabling explicit characterization of design uncertainty. Second, subgroup ATE estimators are asymptotically normal; Third, the final ATE estimator is asymptotically normal for the design-induced estimand, with a variance decomposition that separates uncertainty from learning the enrollment rule and from subgroup effect estimation, justifying the validity of the proposed inference.

We start with presenting the assumptions needed for our theoretical investigation. We write
$O_i=(Y_i,D_i,X_i,S_i)$ for subject $i$'s observed data. Order subjects according to their enrollment time and define
$
\mathcal F_0=\{\varnothing,\Omega\},
\mathcal F_i=\sigma(O_1,\ldots,O_i).
$

\begin{assumption}[Causal framework, sampling, and moments]
\label{ass:causal_sampling_moments_grp}
We assume:
\begin{itemize}
    \item [(a)] The Stable Unit Treatment Value  Assumption (SUTVA) \citep{rubin_randomization_1980} holds:
    There is no interference between units and no hidden versions of the
    treatment. Each subject $i$ has well-defined potential outcomes
    $\{Y_i(1),Y_i(0)\}$, and the observed outcome satisfies
    $Y_i = Y_i(1)D_i+Y_i(0)(1-D_i)$.
    \item[(b)] Within each subgroup $j\in\{1,2\}$ and stage
    $t\in\{1,2\}$, $D_i\in\{0,1\}$, the probability of receiving treatment is strictly between
    0 and 1: $
      0 < \mathbb{P}(D_i=1\mid X_i=j, S_i=t) < 1
      \quad\text{for enrolled subjects.}
    $
\item [(c)] The conditional means and
    variances of the potential outcomes are stable across stages within each
    subgroup: $
    \mathbb{E}\bigl[Y_i(d)\mid X_i=j, S_i=t\bigr]
    = \mathbb{E}\bigl[Y_i(d)\mid X_i=j\bigr],
    \mathbb{V}\bigl[Y_i(d)\mid X_i=j, S_i=t\bigr]
    = \mathbb{V}\bigl[Y_i(d)\mid X_i=j\bigr],$ for all $d\in\{0,1\}$, $j\in\{1,2\}$, and $t\in\{1,2\}$.
\item [(d)]   There exist constants $\eta>0$ and $C<\infty$ such that,
for every $d\in\{0,1\}$, $j\in\{1,2\}$, $t\in\{1,2\}$,
$
\sup_i
\mathbb E\left[|Y_i(d)|^{4+\eta}\middle|\,
  X_i=j, S_i=t, \mathcal F_{i-1}\right]
\le C
$ almost surely
where $\mathcal F_{i-1}$ denotes the history immediately before
subject $i$ is enrolled.
\item [(e)] The potential outcomes have non-vanishing conditional variances, there exists some $v_0>0$ such that $\mathbb{V}[Y_{i}(d)|X_i=j]\ge v_0$ for $d\in\{0,1\}$, $j\in\{1,2\}$.
(f) The asymptotic regime follows $N \rightarrow \infty$ and through out the process $N^{(1)}/N = \kappa$, $\kappa \in (0,1)$. 
\end{itemize}
\end{assumption}

\begin{assumption}[Regularity conditions]\label{ass:regularity}
There exists some $\delta \in (0, \tfrac12)$ such that: (a) 
For all subgroups $j \in \{1,2\}$ and stages $t \in \{1,2\}$,
\(
\delta \le e^{(t)}_j \le 1-\delta .
\)
(b) The Stage 1 (baseline) subgroup prevalence is bounded away from 0 and 1:
\(
\delta < q_1 < 1-\delta \quad (\text{equivalently } \delta<q_2<1-\delta).
\)
\end{assumption}
Assumption~\ref{ass:causal_sampling_moments_grp}(a) is the standard causal assumption ensuring well-defined potential outcomes and no interference. Assumption \ref{ass:causal_sampling_moments_grp}(b) imposes positivity of randomization within each subgroup and stage. Assumption \ref{ass:causal_sampling_moments_grp}(c) is a subgroup-level
transportability condition, requiring stage-invariant conditional means and variances of potential outcomes.
Assumptions \ref{ass:causal_sampling_moments_grp} (d) and (e) impose finite moments and variances bounded away from zero, and 
Assumption \ref{ass:causal_sampling_moments_grp}(f) requires the Stage 1 sample size to be a nonvanishing fraction of the total.  Assumption~\ref{ass:regularity} imposes standard positivity conditions ensuring that the randomization probabilities and baseline subgroup prevalences remain bounded away from the boundary.

\subsection{Theoretical property of the adaptive enrollment probability}

In this section, we study the
theoretical properties of the adaptive enrollment strategy. This is necessary
because the final estimator targets the treatment effect in the trial population induced by the learned enrichment rule.  Under Assumption~\ref{ass:causal_sampling_moments_grp}, Lemma~\ref{lemma:p_star_formal} establishes the properties of the Stage~1–estimated enrollment probability.

\begin{lemma}[Asymptotic linearity of the  optimal proportion]\label{lemma:p_star_formal}
We adopt the following notation for convenience:
$$g(p):=p\log\frac{p}{q_1}+(1-p)\log\frac{1-p}{q_2},\ M(p):=F(p)-\lambda g(p),$$
where $$F(p):= \frac{p\tau_1+(1-p)\tau_2}{\sqrt{pV_1+(1-p)V_2}}.$$
Assume Assumption~\ref{ass:causal_sampling_moments_grp} and Assumption ~\ref{ass:regularity} holds. Further, we need the following condition for penalty parameter $\lambda$.
Write
$
a_\kappa=\kappa q_1,$
$b_\kappa=1-\kappa(1-q_1),$ 
$K_\kappa=\sup_{p\in\mathcal{P}_\kappa}[F''(p)]_+,$
where $[x]_+=\max(x,0)$. When $
(\tau_1-\tau_2)(V_1-V_2)\neq 0,$
let
$
p_c
=
\frac{5V_1\tau_2-6V_2\tau_1+V_2\tau_2}
{(V_1-V_2)(\tau_1-\tau_2)},
$
then $K_\kappa$ has the closed form
$
K_\kappa
=
\max\{
[F''(a_\kappa)]_+,\,
[F''(b_\kappa)]_+,\,
[F''(p_c)]_+
\mathbf{1}\{p_c\in(a_\kappa,b_\kappa)\}
\}.
$
If \(V_1=V_2\), then \(K_\kappa=0\). when \(\tau_1=\tau_2\), the endpoint maximum applies.

\begin{align} \label{eq:lambda_condtion}
\lambda >
\tilde{\lambda}=\max\Bigg\{\frac{K_{\kappa}}{4}, \
\frac{[-F'(\kappa q_1)]_+}{\log\left(\frac{1-\kappa q_1}{\kappa(1-q_1)}\right)},\ 
\frac{[F'\!\big(1-\kappa(1-q_1)\big)]_+}{\log\left(\frac{1-\kappa(1-q_1)}{\kappa q_1}\right)}
\Bigg\}.
\end{align}
This condition is sufficient for strict concavity and interior solution. It remains well
defined when the subgroup effects or variance contributions are equal.
Under the above conditions and as $N\to\infty$, the objective function has a unique maximizer $p_1^*$ in $(\kappa q_1, 1-\kappa(1-q_1))$.  
Let $M_{N^{(1)}}(p) = F(\hat{\theta}^{(1)},p) - \lambda g(p)$.
The Stage~1 optimizer 
\(\hat p_1^{*} = \arg\max_{p_1\in \mathcal P_\kappa} M_{N^{(1)}}(p_1)\) is consistent for
\(p_1^{*}\) and we have:
$
\sqrt{N^{(1)}} \left(\hat p_1^{*}-p_1^{*}\right)
=
\frac{1}{\sqrt{N^{(1)}}}\sum_{i:S_i=1}\psi_{p}^{(1)}(O_i)
+o_p(1),
$
where
\begin{align}\label{eq:psi_p_stage1_impl}
\psi_{p}^{(1)}(O_i)
=
-
\Bigl\{\partial_{pp} M( p_1^*)\Bigr\}^{-1}
\Bigl[\nabla_\theta\,\partial_p F(\theta,p_1^*)\Bigr]^\top
\Phi_i^{(1)}(O_i),
\end{align}
$\Phi_i^{(1)}$ is the (Stage 1) influence-function vector for estimators $\hat\theta^{(1)}=(\hat\tau^{(1)}_1,\hat\tau^{(1)}_2,\hat V^{(1)}_1,\hat V^{(1)}_2)^\top$ of the corresponding parameters $\theta=(\tau_1,\tau_2,V_1,V_2)^\top$.
\end{lemma}

The asymptotic linear result in Lemma \ref{lemma:p_star_formal} is essential, because it ensures that the adaptively
chosen enrollment probability converges to the true optimum, so that the error
introduced by the adaptive step does not prevent the final treatment effect
estimator from achieving standard asymptotic normality. Detailed characterization of Equation \eqref{eq:psi_p_stage1_impl} can be found in Supplementary Materials. The details for the underlying condition \eqref{eq:lambda_condtion} can be found in Supplementary Materials. And the general ideas follow \citep{van_der_vaart_asymptotic_2000,newey_large_1994}.

\subsection{Theoretical properties of the treatment effect estimators}

In this section, we study the theoretical properties of the treatment effect estimators under our proposed design. Theorem~\ref{thm:tau_j_formal} 
establishes asymptotic normality of the subgroup estimators, and Theorem~\ref{thm:psi_formal} provides the main result for the final ATE estimator under the proposed design. We start with
characterizing the asymptotic properties of the subgroup treatment effect estimator.

\begin{theorem}[Asymptotic normality of subgroup estimators]
\label{thm:tau_j_formal}
Assume Assumption~\ref{ass:causal_sampling_moments_grp} and \ref{ass:regularity} hold.  In addition, the parameter $\lambda>\tilde{\lambda}$ is defined above. Then, as $N\to\infty$, we know
the subgroup ATE estimators $\hat\tau_j$ computed using the full data
($N=N^{(1)}+N^{(2)}$) are consistent and asymptotically linear:
\begin{align}
    \sqrt{N}\,(\hat\tau_j - \tau_j)
\;= \frac{1}{\sqrt{N}}\sum_{i=1}^{N}\psi_j(O_i) +o_p{(1)}.
\end{align}

And asymptotic normality with
\begin{align}
    \sqrt{N}\,(\hat\tau_j - \tau_j)
\;\rightsquigarrow\; N(0, \mathbb{V}\{\psi_j(O_i)\}),
\end{align}

where 
\begin{align}
    \psi_j(O_i)&=\frac{N\,\mathbf{1}(X_i=j)D_i}{\sum_{i=1}^N\mathbf{1}(X_i=j)D_i}\,\bigl(Y_i-\mu_j(1)\bigr)\\\notag
&- \frac{N\,\mathbf{1}(X_i=j)(1-D_i)}{\sum_{i=1}^N\mathbf{1}(X_i=j)(1-D_i)}\,\bigl(Y_i-\mu_j(0)\bigr).
\end{align}

\end{theorem}

Theorem~\ref{thm:tau_j_formal} establishes asymptotic normality of the subgroup treatment effect estimators Theorem~\ref{thm:tau_j_formal} is an important component for establishing the theoretical result for the ATE estimator, since our ATE estimator is $
\hat\tau \;=\; \hat p_1^{*}\,\hat\tau_1 + (1-\hat p_1^{*})\,\hat\tau_2$.
Thus, the asymptotic
behavior of \(\hat\tau\) depends on both the estimated optimal enrichment probability
\(\hat p_1^{*}\) and the subgroup estimators \(\hat\tau_j\). Together with Lemma \ref{lemma:p_star_formal}, we establish our main result for the ATE estimator in Theorem \ref{thm:psi_formal}. 

\begin{theorem}[Asymptotic normality of the overall ATE estimator]
\label{thm:psi_formal}
Under Assumptions~\ref{ass:causal_sampling_moments_grp} and \ref{ass:regularity}, consider
\begin{align}
    \hat\tau
= \hat p_1^{*}\,\hat\tau_1 + (1-\hat p_1^{*})\,\hat\tau_2
\end{align}
as an estimator of the optimal ATE $\tau(p_1^{*})$. In addition, the parameter $\lambda>\tilde{\lambda}$ as defined above. Then, as $N\to\infty$, we have.
\begin{align}
    \sqrt{N}\,\bigl(\hat\tau - \tau(p_1^{*})\bigr)
=
\frac{1}{\sqrt{N}}\sum_{i=1}^N \psi_\tau(O_i) + o_{\mathbb{P}}(1).
\end{align}

Then
\begin{align}
    \sqrt{N}\,\bigl(\hat\tau - \tau(p_1^{*})\bigr)
\;\rightsquigarrow\; N(0, \mathbb{V}\{\psi_\tau(O_i)\}),
\end{align}
where $\psi_\tau(O_i)$ is decomposed from functional delta method.
\begin{align}
    \psi_\tau(O_i)
=
(\tau_1-\tau_2)\,\psi_{p}(O_i)
+ p_1^{*}\,\psi_{1}(O_i)
+ (1-p_1^{*})\,\psi_{2}(O_i),
\end{align}
and we need to scale the influence function from Lemma \ref{lemma:p_star_formal}
\begin{align}\label{eq:psi_p_scaling}
\psi_{p}(O_i)
:=
\frac{\mathbf{1}(S_i=1)}{\kappa}\,\psi_{p}^{(1)}(O_i).
\end{align}
\end{theorem}

Theorem~\ref{thm:psi_formal} establishes asymptotic normality of the ATE estimator under the proposed design.
Because the ATE estimator is a
weighted average of subgroup estimators with a weight learned
adaptively from Stage 1, the estimand is design-induced, and therefore, valid inference must account for uncertainty in both subgroup effects and the estimated optimal enrollment probability. Establishing this theoretical result combines $\sqrt{N^{(1)}}$-asymptotic linearity for the interim
optimizer $\hat p_1^*$
(Lemma~\ref{lemma:p_star_formal}) and 
$\sqrt{N}$-asymptotic
normality for the subgroup estimators $\hat\tau_j$ (Theorem~\ref{thm:tau_j_formal}). 
Since the inference target is the smooth map
$g(p,\tau_1,\tau_2)=p\tau_1+(1-p)\tau_2$, the functional delta method yields an influence-function
representation for $\hat\tau$, which is composed of two components. $(\tau_1-\tau_2)\psi_p(O_i)$ quantifies the variability contributed by learning the optimal enrollment probability from interim data.
Its magnitude is scaled by heterogeneity: if $\tau_1$ is close to $\tau_2$, then small errors in the
mixing weight have little impact on the overall ATE. $p_1^*\psi_1(O_i)+(1-p_1^*)\psi_2(O_i)$ captures the influence function 
from estimating subgroup treatment effects. 
In the two theorems, the $\psi$-terms are finite-sample
stabilized summands. Because their denominators use full-sample
subgroup-arm counts, the martingale central limit theorem is applied
to the underlying centered numerator sums, while the denominators are
handled by consistency and Slutsky's theorem.

\section{Simulation studies}
\label{sec:sim}

In this section, we evaluate the finite-sample operating characteristics of
the proposed adaptive enrichment design. We summarize the main takeaways
as follows. First, under the strong null, the proposed procedure maintains
rejection probabilities close to the nominal level across different values of the regularization parameter $\lambda$. Second, under
treatment-effect heterogeneity, $\lambda$ governs an enrichment-stability
trade-off. Third, the direct bias and design-aligned MSE
analyses clarify the mechanism behind this trade-off: regularization can
substantially reduce selection-induced instability and finite-sample
estimation error when the enrichment decision is difficult to learn, but it
does not provide uniform MSE improvement when the unpenalized decision is
already stable.

\subsection{Simulation setup}\label{subsec:simulation setup}

We consider a two-stage adaptive enrichment trial with two prespecified
subgroups. Let $X_i\in\{1,2\}$ denote the subgroup indicator, with $q_1=0.5
$
The two stages have equal planned sample sizes, so that
$
\kappa=\frac{N^{(1)}}{N}=0.5,
N^{(1)}=N^{(2)}=\frac{N}{2}.
$
Stage~1 enrollment follows the baseline subgroup prevalence $q_1=0.5$.
The Stage~2 enrollment distribution is determined by the procedure being
evaluated.

Within each subgroup and stage, subjects are independently randomized to
treatment with probability 0.5.
The potential outcomes are generated according to
$
Y_i(0)\mid X_i=j\sim N(0,1),
Y_i(1)\mid X_i=j\sim N(\tau_j,1), j\in\{1,2\},
$
and the observed outcome is
$
Y_i=D_iY_i(1)+(1-D_i)Y_i(0).
$
Because Stage~1 enrolls a fraction $\kappa=0.5$ of the total sample at
prevalence $q_1=0.5$, the feasible cumulative subgroup-1 proportion is
restricted to
$
\mathcal P_{\kappa}
=
\left[
\kappa q_1,\,
1-\kappa(1-q_1)
\right]
=
[0.25,0.75].
$

We compare the following four design and analysis procedures. (i)  ``Nonadaptive design" continues Stage~2 enrollment at the baseline
prevalence,
$
p_1^{(2)}=q_1=0.5,
$
and targets the baseline-population treatment effect
$
\tau(q_1)=q_1\tau_1+(1-q_1)\tau_2.
$
Its one-sided null hypothesis is
$
H_{0,\mathrm{NA}}:\tau(q_1)\leq 0.
$
(ii) ``Proposed design" uses Stage~1 data to solve the empirical
regularized optimization problem in Section~3. Let
$\widehat p_{1,\lambda}^{\star}$ denote the resulting estimated cumulative
subgroup-1 proportion under penalty $\lambda$. 
The oracle design-induced target is
$
\tau_\lambda
=
\tau\!\left(p_{1,\lambda}^{\star}\right)
=
p_{1,\lambda}^{\star}\tau_1
+
\left(1-p_{1,\lambda}^{\star}\right)\tau_2
$. The corresponding one-sided null hypothesis is
$
H_{0,\lambda}:
\tau\left(p_{1,\lambda}^{\star}\right)\leq 0.
$
We consider
$
\lambda\in\{0,0.21,0.50,1\}.
$
The value $\lambda=0$ is included as an unpenalized finite-sample
benchmark. Because the unpenalized objective may have a boundary or
nonunique optimizer, the $\lambda=0$ results are reported as empirical
operating characteristics and are not intended to verify the
interior-optimizer asymptotic results in Section~5. (iii) The enriched-grid Holm procedure, denoted ``Enr.\ Holm," considers
the prespecified family of enriched populations
$
p\in\{0.25,0.50,0.75\}.
$ Holm's procedure is applied to the associated one-sided
hypothesis family, and the table reports rejection of the adjusted test for
the selected enriched population.
(iv) The \emph{pick-the-winner procedure}, denoted ``PTW selected,"
selects the subgroup with the larger Stage~1 treatment-effect estimate,
enrolls only that subgroup during Stage~2, and tests the one-sided null
hypothesis for the subgroup selected at Stage~1. For the four procedures, their
rejection probabilities describe the operating characteristics
of the corresponding design and analysis procedures.

The rejection-probability study uses $N=1000$ subjects,
$R=2000$ Monte Carlo replicates, and $B=200$ full-algorithm parametric
bootstrap draws for each simulated trial. We consider the strong-null
setting
$
(\tau_1,\tau_2)=(0,0),
$
a small-gap setting
$
(\tau_1,\tau_2)=(0.10,0.15),
$
and a clear-gap setting
$
(\tau_1,\tau_2)=(0,0.40).
$
The global-null scaled-bias experiment uses $N=2000$ and $R=2000$
Monte Carlo replicates. Finally, the enrollment-stability and
design-aligned MSE experiment uses $N=2000$ and $R=5000$ Monte Carlo
replicates under the small-gap and clear-gap settings.

\subsection{Rejection probabilities across regularization levels}\label{subsec:simulation results:power comparison}

Table~\ref{tab:type1} reports one-sided rejection
probabilities for the proposed and benchmark procedures. Under the strong
null, every treatment-effect target considered in the table is zero, so the
reported rejection probabilities represent Type I error. Under the
small-gap and clear-gap settings, the entries represent power against the
corresponding method-specific alternatives.

Under the strong null, the rejection probabilities are close to the nominal level. The Enr.\ Holm procedure is conservative in this setting, whereas the PTW selected procedure has rejection probability
$0.071$. The latter two columns use different hypothesis structures and
exhibit different null calibration in this experiment. Their nonnull
rejection probabilities should therefore be regarded as contextual
operating characteristics rather than as equal-size power comparisons with
the proposed design.

In the small-gap setting, the unpenalized rule
$\lambda=0$ permits the most aggressive enrichment toward subgroup~2 and
therefore produces the largest oracle design-induced treatment-effect
target among the proposed procedures. Nevertheless, its rejection
probability is below those of the three positive-penalty
procedures. The positive penalties stabilize the Stage~1 estimate of the
enrollment proportion, and this reduction in decision variability offsets
the modest reduction in the design-induced target. This setting demonstrates that more aggressive enrichment
does not necessarily produce greater power when the corresponding
enrollment decision is difficult to estimate reliably.
In the clear-gap setting, subgroup 2 is
substantially more responsive than subgroup~1. The unpenalized boundary
decision is much easier to learn from Stage~1 data. In this
case, the proposed rejection probability decreases gradually as the
regularization becomes stronger.  

Overall, Table~\ref{tab:type1} shows that the effect of
$\lambda$ is scenario dependent. When the subgroup-effect gap is small,
moderate regularization can improve power by stabilizing the learned
enrollment mixture. When the subgroup-effect gap is clear, a smaller value
of $\lambda$ can yield greater power because the aggressive enrichment
decision is already stable. Thus, the simulations do not identify a
universally optimal value of $\lambda$.

\begin{table}[t]

\centering
\caption{\small One-sided rejection probabilities at $\alpha=0.05$. The strong-null row
reports Type I error. The other rows report power under the corresponding
method-specific alternatives. The maximum Monte Carlo standard error is
0.011.}
\label{tab:type1}
\setlength{\tabcolsep}{3.2pt}
\begin{adjustbox}{max width=\textwidth}
\begin{tabular}{lccccccc}
\toprule
Scenario $(\tau_1,\tau_2)$
& Nonadaptive
& Prop.\ $0$
& Prop.\ $0.21$
& Prop.\ $0.50$
& Prop.\ $1$
& Enr.\ Holm
& PTW selected \\
\midrule
Strong null $(0,0)$       & .042 & .039 & .045 & .048 & .053 & .012 & .071 \\
Small gap $(0.10,0.15)$   & .623 & .600 & .627 & .640 & .630 & .374 & .641 \\
Clear gap $(0,0.40)$      & .927 & .990 & .988 & .980 & .968 & .978 & .988 \\
\bottomrule
\end{tabular}
\end{adjustbox}

\end{table}

\subsection{Effect of penalty term}\label{subsec:simulation results:effect of penalization term}

First, we examine selection-induced bias under the global null. Because
$\tau_1=\tau_2=0$ in this experiment, the trial-population ATE equals
zero for every possible enrollment mixture. Consequently, changing
$\lambda$ does not change the estimand, and the design-aligned and
fixed-reference definitions of bias coincide. We provide a figure reporting $\sqrt{N}$-scaled bias as a
function of the regularization strength $\lambda$ in Supplementary Materials.
The PTW design exhibits non-negligible scaled bias. Under
the proposed design, the scaled bias decreases as the regularization
strength increases, as the KL penalty anchors the
enrollment decision toward the baseline subgroup prevalence. 

Second, we study in detail the effect of regularization on bias, enrollment stability, and the mean-squared error (MSE). 
Table~\ref{tab:lambda-mse} reports design-aligned error and the target
$\tau_\lambda$ separately, rather than combining target shift with
estimation error in the reported RMSE. Note that the $\lambda=0$ rows are included as
finite-sample unpenalized benchmarks.

\begin{table}[htbp]

\centering
 \caption{\small Monte Carlo operating characteristics relative to each
design-aligned target. }
\label{tab:lambda-mse}

\setlength{\tabcolsep}{3pt}
\begin{adjustbox}{max width=\linewidth}
\begin{tabular}{lcccccccc}
\toprule
Scenario $(\tau_1,\tau_2)$
& $\lambda$
& \shortstack{Optimal prevalence\\$p^\star_\lambda$}
& \shortstack{Monte Carlo mean\\$\mathbb{E}(\widehat p^\star_\lambda)$}
& \shortstack{Target\\$\tau(p^\star_\lambda)$}
& \shortstack{SD of\\$\widehat p^\star_\lambda$}
& \shortstack{$\sqrt{N}$\\Bias}
& \shortstack{$\sqrt{N}$\\Empirical SD}
& \shortstack{$\sqrt{N}$ RMSE\\(MCSE)}
\\
\midrule
Small gap $(0.10,0.15)$
& 0
& 0.250
& 0.4245
& 0.1375
& 0.2384
& $-0.418$
& 2.057
& 2.099 (0.021)
\\
& 0.21
& 0.470
& 0.4704
& 0.1265
& 0.0754
& $-0.003$
& 2.015
& 2.015 (0.020)
\\
& 0.50
& 0.488
& 0.4873
& 0.1256
& 0.0324
& 0.028
& 1.951
& 1.951 (0.019)
\\
& 1
& 0.494
& 0.4936
& 0.1253
& 0.0162
& 0.015
& 1.999
& 1.999 (0.020)
\\
\addlinespace
Clear gap $(0,0.40)$
& 0
& 0.250
& 0.2507
& 0.3000
& 0.0187
& $-0.026$
& 2.003
& 2.003 (0.022)
\\
& 0.21
& 0.278
& 0.2932
& 0.2886
& 0.0485
& $-0.237$
& 2.240
& 2.252 (0.022)
\\
& 0.50
& 0.401
& 0.4014
& 0.2395
& 0.0311
& 0.020
& 2.001
& 2.001 (0.019)
\\
& 1
& 0.450
& 0.4501
& 0.2199
& 0.0160
& 0.032
& 2.004
& 2.004 (0.020)
\\
\bottomrule
\end{tabular}
\end{adjustbox}

\end{table}

In the small-gap setting, the unpenalized boundary decision is difficult to learn from the
Stage~1 data. Positive regularization substantially stabilizes the
estimated enrollment proportion, sharply reducing
$\operatorname{SD}(\widehat p_{1,\lambda}^{\star})$. It also lowers the
design-aligned $\sqrt N$-scaled RMSE throughout the positive-penalty
grid. In the clear-gap setting, the
unpenalized rule already selects the oracle boundary almost
deterministically. 

For $\lambda=0.50$ and $\lambda=1$, the
design-aligned RMSE is again approximately equal to that of the
unpenalized rule. However, the target decreases as $\lambda$ increases. Under the common
reference target $\tau_0$, these changes would enter the error criterion
as regularization-induced target shift, even though the estimators remain
accurate for their respective design-induced targets. In sum, Table~\ref{tab:lambda-mse} show that regularization can reduce selection-induced instability
when the subgroup effects are difficult to distinguish. We provide additional theoretical insights in Supplementary Materials Section 4.

\section{Discussion}

We develop a unified, optimization-based adaptive enrichment framework that casts interim enrollment adaptation as a regularized power-maximization problem, making the power and distribution drift trade-off explicit and pre-specifiable. There are several directions that warrant future exploration. 
First, although we adopt a KL divergence penalty, a systematic comparison with alternative discrepancy measures and constraints is an important direction for future work, as these alternatives may offer different robustness properties or better reflect application-specific notions of distributional drift.
Second, extending the framework to incorporate other design objectives, such as operational constraints, would enhance practical applicability in practical enrichment trials.

\bibliography{references}

@article{ruan2026hero,
  title={HERO: Improving the Reliability and Sensitivity of Generative Model Evaluation Using Historical Data},
  author={Ruan, Xinrui and Zhao, Zhenyu and Wei, Waverly and Zhang, Yueshan and Zheng, Zeyu and Huang, Sui and Wang, Jingshen},
  journal={arXiv preprint arXiv:2606.29784},
  year={2026}
}

@article{csiszar_i-divergence_1975,
	title = {I-{Divergence} {Geometry} of {Probability} {Distributions} and {Minimization} {Problems}},
	volume = {3},
	issn = {0091-1798},
	url = {https://www.jstor.org/stable/2959270},
	number = {1},
	urldate = {2025-12-28},
	journal = {The Annals of Probability},
	author = {Csiszár, I.},
	year = {1975},
	note = {Publisher: Institute of Mathematical Statistics},
	pages = {146--158},
}

@article{newey_large_1994,
	title = {Large sample estimation and hypothesis testing},
	volume = {4},
	url = {https://www.sciencedirect.com/science/article/pii/S1573441205800054},
	urldate = {2025-12-28},
	journal = {Handbook of econometrics},
	author = {Newey, Whitney K. and McFadden, Daniel},
	year = {1994},
	note = {Publisher: Elsevier},
	pages = {2111--2245},
}

@article{rosenblum_optimizing_2011,
	title = {Optimizing randomized trial designs to distinguish which subpopulations benefit from treatment},
	volume = {98},
	issn = {0006-3444, 1464-3510},
	url = {https://academic.oup.com/biomet/article-lookup/doi/10.1093/biomet/asr055},
	doi = {10.1093/biomet/asr055},
	language = {en},
	number = {4},
	urldate = {2024-07-22},
	journal = {Biometrika},
	author = {Rosenblum, M. and Van Der Laan, M. J.},
	month = dec,
	year = {2011},
	pages = {845--860},
}

@article{marcus_closed_1976,
	title = {On {Closed} {Testing} {Procedures} with {Special} {Reference} to {Ordered} {Analysis} of {Variance}},
	volume = {63},
	issn = {0006-3444},
	url = {https://www.jstor.org/stable/2335748},
	doi = {10.2307/2335748},
	number = {3},
	urldate = {2025-12-28},
	journal = {Biometrika},
	author = {Marcus, Ruth and Peritz, Eric and Gabriel, K. R.},
	year = {1976},
	note = {Publisher: [Oxford University Press, Biometrika Trust]},
	pages = {655--660},
}

@article{fda_enrichment_2019,
	title = {Enrichment {Strategies} for {Clinical} {Trials} to {Support} {Determination} of {Effectiveness} of {Human} {Drugs} and {Biological} {Products} {Guidance} for {Industry}},
	language = {en},
	author = {FDA},
	year = {2019},
}

@article{zhang_treatment_2018,
	title = {Treatment evaluation for a data-driven subgroup in adaptive enrichment designs of clinical trials},
	volume = {37},
	copyright = {Copyright © 2017 John Wiley \& Sons, Ltd.},
	issn = {1097-0258},
	url = {https://onlinelibrary.wiley.com/doi/abs/10.1002/sim.7497},
	doi = {10.1002/sim.7497},
	language = {en},
	number = {1},
	urldate = {2025-10-16},
	journal = {Statistics in Medicine},
	author = {Zhang, Zhiwei and Chen, Ruizhe and Soon, Guoxing and Zhang, Hui},
	year = {2018},
	note = {\_eprint: https://onlinelibrary.wiley.com/doi/pdf/10.1002/sim.7497
TLDR: A 2‐stage AED is proposed which does not require predefined subgroups but requires a prespecified algorithm for choosing a subgroup on the basis of baseline covariate information and is evaluated and compared in a simulation study mimicking actual clinical trials of human immunodeficiency virus infection.},
	pages = {1--11},
}

@article{stallard_adaptive_2014,
	title = {Adaptive designs for confirmatory clinical trials with subgroup selection},
	volume = {24},
	issn = {1520-5711},
	doi = {10.1080/10543406.2013.857238},
	language = {eng},
	number = {1},
	journal = {Journal of Biopharmaceutical Statistics},
	author = {Stallard, Nigel and Hamborg, Thomas and Parsons, Nicholas and Friede, Tim},
	year = {2014},
	pmid = {24392984},
	pages = {168--187},
}

@article{simon_adaptive_2013,
	title = {Adaptive enrichment designs for clinical trials},
	volume = {14},
	issn = {1465-4644, 1468-4357},
	url = {https://academic.oup.com/biostatistics/article-lookup/doi/10.1093/biostatistics/kxt010},
	doi = {10.1093/biostatistics/kxt010},
	language = {en},
	number = {4},
	urldate = {2024-07-22},
	journal = {Biostatistics},
	author = {Simon, N. and Simon, R.},
	month = sep,
	year = {2013},
	note = {TLDR: A class of adaptive enrichment designs that allow the eligibility criteria of a trial to be adaptively updated during the trial, restricting entry to patients likely to benefit from the new treatment are proposed.},
	pages = {613--625},
}

@article{frieri_design_2023,
	title = {Design {Considerations} for {Two}-{Stage} {Enrichment} {Clinical} {Trials}},
	volume = {79},
	copyright = {https://academic.oup.com/journals/pages/open\_access/funder\_policies/chorus/standard\_publication\_model},
	issn = {0006-341X, 1541-0420},
	url = {https://academic.oup.com/biometrics/article/79/3/2565-2576/7513870},
	doi = {10.1111/biom.13805},
	language = {en},
	number = {3},
	urldate = {2024-07-22},
	journal = {Biometrics},
	author = {Frieri, Rosamarie and Rosenberger, William Fisher and Flournoy, Nancy and Lin, Zhantao},
	month = sep,
	year = {2023},
	note = {TLDR: A two‐stage enrichment design is described, in which the first stage is designed to efficiently estimate a threshold and the second stage is a “phase III‐like” trial on the enriched population, described by a bivariate normal model.},
	pages = {2565--2576},
}

@article{baldi_antognini_new_2023,
	title = {New insights into adaptive enrichment designs},
	volume = {64},
	issn = {0932-5026, 1613-9798},
	url = {https://link.springer.com/10.1007/s00362-023-01433-0},
	doi = {10.1007/s00362-023-01433-0},
	language = {en},
	number = {4},
	urldate = {2024-07-22},
	journal = {Statistical Papers},
	author = {Baldi Antognini, Alessandro and Frieri, Rosamarie and Zagoraiou, Maroussa},
	month = aug,
	year = {2023},
	note = {TLDR: This review is dedicated to adaptive enrichment studies with a focus on design aspects, and discusses the multiple aspects involved in adaptive enrichment designs that contribute to their advantages and disadvantages.},
	pages = {1305--1328},
}

@article{simon_inference_2017,
	title = {Inference for multimarker adaptive enrichment trials},
	volume = {36},
	copyright = {http://onlinelibrary.wiley.com/termsAndConditions\#vor},
	issn = {0277-6715, 1097-0258},
	url = {https://onlinelibrary.wiley.com/doi/10.1002/sim.7422},
	doi = {10.1002/sim.7422},
	language = {en},
	number = {26},
	urldate = {2024-07-22},
	journal = {Statistics in Medicine},
	author = {Simon, Richard and Simon, Noah},
	month = nov,
	year = {2017},
	pages = {4083--4093},
}

@article{rosenblum_optimal_2020,
	title = {Optimal, {Two}-{Stage}, {Adaptive} {Enrichment} {Designs} for {Randomized} {Trials}, using {Sparse} {Linear} {Programming}},
	volume = {82},
	copyright = {https://academic.oup.com/journals/pages/open\_access/funder\_policies/chorus/standard\_publication\_model},
	issn = {1369-7412, 1467-9868},
	url = {https://academic.oup.com/jrsssb/article/82/3/749/7056063},
	doi = {10.1111/rssb.12366},
	language = {en},
	number = {3},
	urldate = {2024-07-22},
	journal = {Journal of the Royal Statistical Society Series B: Statistical Methodology},
	author = {Rosenblum, Michael and Fang, Ethan X. and Liu, Han},
	month = jul,
	year = {2020},
	note = {TLDR: The key to the approach is a novel, discrete representation of this optimization problem as a sparse linear program, which is large but computationally feasible to solve by using modern optimization techniques.},
	pages = {749--772},
}

@article{rosenblum_multiple_2016,
	title = {Multiple testing procedures for adaptive enrichment designs: combining group sequential and reallocation approaches},
	volume = {17},
	issn = {1465-4644},
	shorttitle = {Multiple testing procedures for adaptive enrichment designs},
	url = {https://doi.org/10.1093/biostatistics/kxw014},
	doi = {10.1093/biostatistics/kxw014},
	number = {4},
	urldate = {2024-01-24},
	journal = {Biostatistics},
	author = {Rosenblum, Michael and Qian, Tianchen and Du, Yu and Qiu, Huitong and Fisher, Aaron},
	month = oct,
	year = {2016},
	note = {TLDR: A new class of multiple testing procedures tailored to adaptive enrichment designs is proposed, proved to have power greater than or equal to several existing methods, and to strongly control the familywise Type I error rate when statistics are normally distributed.},
	pages = {650--662},
}

@article{magnusson_group_2013,
	title = {Group sequential enrichment design incorporating subgroup selection},
	volume = {32},
	copyright = {http://onlinelibrary.wiley.com/termsAndConditions\#vor},
	issn = {0277-6715, 1097-0258},
	url = {https://onlinelibrary.wiley.com/doi/10.1002/sim.5738},
	doi = {10.1002/sim.5738},
	language = {en},
	number = {16},
	urldate = {2024-07-22},
	journal = {Statistics in Medicine},
	author = {Magnusson, Baldur P. and Turnbull, Bruce W.},
	month = jul,
	year = {2013},
	note = {TLDR: Numerical results show that the adaptive enrichment group sequential procedure has high power to detect subgroup‐specific effects and the use of multiple interim analysis points can lead to substantial sample size savings.},
	pages = {2695--2714},
}

@book{van_der_vaart_asymptotic_2000,
	title = {Asymptotic statistics},
	volume = {3},
	url = {https://books.google.com/books?hl=en&lr=&id=Ocg2AAAAQBAJ&oi=fnd&pg=PR13&dq=info:x8RE2hPsLVEJ:scholar.google.com&ots=Rp5Xu0KP6K&sig=KRNSyUQK2cTOsOCZ-Q_ly6wJ5oc},
	urldate = {2025-12-17},
	publisher = {Cambridge university press},
	author = {Van der Vaart, Aad W.},
	year = {2000},
}

@article{neyman_two_1934,
	title = {On the {Two} {Different} {Aspects} of the {Representative} {Method}: {The} {Method} of {Stratified} {Sampling} and the {Method} of {Purposive} {Selection}},
	volume = {97},
	issn = {0952-8385},
	shorttitle = {On the {Two} {Different} {Aspects} of the {Representative} {Method}},
	url = {https://www.jstor.org/stable/2342192},
	doi = {10.2307/2342192},
	number = {4},
	urldate = {2025-10-03},
	journal = {Journal of the Royal Statistical Society},
	author = {Neyman, Jerzy},
	year = {1934},
	note = {Publisher: [Wiley, Royal Statistical Society]},
	pages = {558--625},
}

@misc{montesuma_recent_2024,
	title = {Recent {Advances} in {Optimal} {Transport} for {Machine} {Learning}},
	url = {http://arxiv.org/abs/2306.16156},
	doi = {10.48550/arXiv.2306.16156},
	urldate = {2025-04-04},
	publisher = {arXiv},
	author = {Montesuma, Eduardo Fernandes and Mboula, Fred Ngolè and Souloumiac, Antoine},
	month = aug,
	year = {2024},
	note = {arXiv:2306.16156 [cs]},
}

@article{lin_inference_2021,
	title = {Inference for a two-stage enrichment design},
	volume = {49},
	issn = {0090-5364, 2168-8966},
	url = {https://projecteuclid.org/journals/annals-of-statistics/volume-49/issue-5/Inference-for-a-two-stage-enrichment-design/10.1214/21-AOS2051.full},
	doi = {10.1214/21-AOS2051},
	number = {5},
	urldate = {2024-01-19},
	journal = {The Annals of Statistics},
	author = {Lin, Zhantao and Flournoy, Nancy and Rosenberger, William F.},
	month = oct,
	year = {2021},
	note = {Publisher: Institute of Mathematical Statistics},
	pages = {2697--2720},
}

@article{stallard_adaptive_2023,
	title = {Adaptive enrichment designs with a continuous biomarker},
	volume = {79},
	issn = {0006-341X, 1541-0420},
	url = {https://onlinelibrary.wiley.com/doi/10.1111/biom.13644},
	doi = {10.1111/biom.13644},
	language = {en},
	number = {1},
	urldate = {2023-10-09},
	journal = {Biometrics},
	author = {Stallard, Nigel},
	month = mar,
	year = {2023},
	pages = {9--19},
}

@article{rubin_estimating_1974,
	title = {Estimating causal effects of treatments in randomized and nonrandomized studies.},
	volume = {66},
	issn = {1939-2176, 0022-0663},
	url = {http://doi.apa.org/getdoi.cfm?doi=10.1037/h0037350},
	doi = {10.1037/h0037350},
	language = {en},
	number = {5},
	urldate = {2022-12-17},
	journal = {Journal of Educational Psychology},
	author = {Rubin, Donald B.},
	month = oct,
	year = {1974},
	pages = {688--701},
}

@article{rubin_randomization_1980,
	title = {Randomization {Analysis} of {Experimental} {Data}: {The} {Fisher} {Randomization} {Test} {Comment}},
	volume = {75},
	issn = {0162-1459},
	shorttitle = {Randomization {Analysis} of {Experimental} {Data}},
	url = {https://www.jstor.org/stable/2287653},
	doi = {10.2307/2287653},
	number = {371},
	urldate = {2022-09-26},
	journal = {Journal of the American Statistical Association},
	author = {Rubin, Donald B.},
	year = {1980},
	note = {Publisher: [American Statistical Association, Taylor \& Francis, Ltd.]},
	pages = {591--593},
}

@article{splawa-neyman_application_1923,
	title = {On the {Application} of {Probability} {Theory} to {Agricultural} {Experiments}. {Essay} on {Principles}. {Section} 9},
	volume = {5},
	issn = {0883-4237, 2168-8745},
	url = {https://projecteuclid.org/journals/statistical-science/volume-5/issue-4/On-the-Application-of-Probability-Theory-to-Agricultural-Experiments-Essay/10.1214/ss/1177012031.full},
	doi = {10.1214/ss/1177012031},
	number = {4},
	urldate = {2022-05-15},
	journal = {Statistical Science},
	author = {Splawa-Neyman, Jerzy and Dabrowska, D. M. and Speed, T. P.},
	year = {1923},
	note = {Publisher: Institute of Mathematical Statistics},
	pages = {465--472},
}

\end{document}